# *NuSTAR* spectral analysis of two bright Seyfert 1 galaxies: MCG +8-11-11 and NGC 6814

A. Tortosa[1]⋆, S. Bianchi[1], A. Marinucci[1], G. Matt[1], R. Middei[1], E. Piconcelli[2],
L. W. Brenneman[3], M. Cappi[4], M. Dadina[4], A. De Rosa [5], P. O. Petrucci[6], F. Ursini[4],
D. J. Walton[7]

[1]*Dipartimento di Matematica e Fisica, Universitá degli Studi Roma Tre, via della Vasca Navale 84, 00146 Roma, Italy.*
[2]*Osservatorio Astronomico di Roma (INAF), via Frascati 33, I-00040 Monte Porzio Catone (Roma), Italy.*
[3]*Smithsonian Astrophysical Observatory, 60 Garden St., MS-4, Cambridge, MA 02138.*
[4]*INAF-IASF Bologna, Via Gobetti 101, I-40129 Bologna, Italy.*
[5]*INAF/Istituto di Astrofisica e Planetologia Spaziali, via Fosso del Cavaliere, 00133 Roma, Italy.*
[6]*Univ. Grenoble Alpes, CNRS, IPAG, F-38000 Grenoble, France.*
[7]*Institute of Astronomy, University of Cambridge, Madingley Road, Cambridge CB3 0HA, UK*

**ABSTRACT**

We report on the *NuSTAR* observations of two bright Seyfert 1 galaxies, namely MCG +8-11-11 (100 ks) and NGC 6814 (150 ks). The main goal of these observations was to investigate the Comptonization mechanisms acting in the innermost regions of AGN which are believed to be responsible for the UV/X-ray emission. The spectroscopic analysis of the *NuSTAR* spectra of these two sources revealed that although they had different properties overall (black hole masses, luminosity and Eddington ratios) they had very similar coronal properties. Both presented a power law spectrum with a high-energy cutoff at ∼ 150−200 keV, a relativistically broadened Fe Kα line and the associated disk reflection component, plus a narrow iron line likely emitted in Compton thin and distant matter. The intrinsic continuum was well described by Comptonization models that show for MCG +8-11-11 a temperature of the coronal plasma of $kT_e$ ∼ 60 keV and an extrapolated optical depth $\tau$=1.8; for NGC 6814 the coronal temperature was $kT_e$ ∼ 45 keV with an extrapolated optical depth of $\tau$=2.5. We compare and discuss these values to some most common Comptonization models which aim at explaining the energy production and stability of coronae in active galactic nuclei.

## 1 INTRODUCTION

The primary X-ray emission of Active Galactic Nuclei (AGN), according to the standard paradigm, is due to thermal Comptonization of the soft disk photons in a hot, optically thin plasma (the so-called corona) located above the accretion disk (e.g., Shapiro et al. 1976, Sunyaev & Titarchuk 1980, Haardt & Matt 1993, Haardt & Maraschi 1991, 1993, Haardt et al. 1994). The spectral shape of this component is, in the first approximation, a power law with a cutoff at high energy. The exact shape of the cutoff depends on the properties of the corona; it is customary to approximate it with an exponential function, at least for a preliminary, phenomenological analysis. The cutoff energy is then expected to be 2-3 times the plasma electron temperature, while the power law index is related to both the temperature and the optical depth of the corona (Petrucci et al. 2000, 2001). The primary emission can illuminate the accretion disk itself or more distant material, like the molecular torus, producing a reflection spectrum (Matt, Perola & Piro 1991, George & Fabian 1991). This spectrum is characterized by fluorescent lines, the most prominent being the Fe Kα line at 6.4 keV (e.g., Matt et al. 1997, Nandra & Pounds 1994, Reynolds & Nowak 2003), and by a broad continuum peaking at ∼ 20 − 30 keV, known as the Compton hump. When the reflection arises from the inner regions of the accretion disc, it is affected by the strong gravitational field of the black hole. The result is an Fe Kα emission line with the shape modified into a skewed and asymmetric profile (Fabian et al. 1989, 2002, 2009, Tanaka et al. 1995). Since the relativistic lines originate within a few gravitational radii from the central object, the analysis of these features represents a powerful probe of the innermost region of the AGN.

⋆ E-mail: tortosa@fis.uniroma3.it

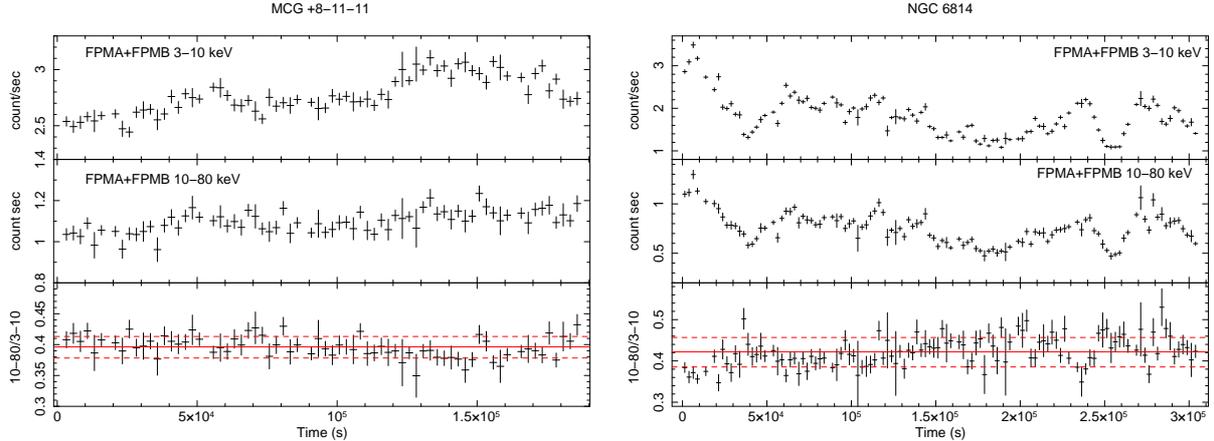

**Figure 1.** The *NuSTAR* FPMA+B light curves in the 3-10 keV (top panels), and in the 10-80 keV energy band (middle panels) are shown, for MCG +8-11-11(left panel) and for NGC 6814 (right panel). The ratio between 3-10 keV and 10-80 keV *NuSTAR* light curves is shown in the bottom panels for bottom sources; the red solid and dashed lines indicate the mean and standard deviation, respectively.

With the superior sensitivity of *NuSTAR* (*Nuclear Spectroscopic Telescope Array*: Harrison et al. 2013) above 10 keV it is possible to separate the primary and reflected continua and measure the coronal parameters, breaking the degeneracy occurring when the high energy cutoff can not be measured. Indeed, a number of high-energy cutoff measurements in local Seyfert galaxies, on a wide range of Eddington ratios, have been already obtained (Fabian et al. 2015, 2017, Marinucci et al. 2016 and references therein), to investigate the Comptonization mechanisms acting in the innermost regions of AGN and which are believed to be responsible for the X-ray emission. However more, and more precise measurements, are needed in order to put these studies on a more firm statistical ground and constrain the coronal parameters. NGC 6814 and MCG +8-11-11 are two bright radio quiet unobscured Seyfert 1 galaxies; they are ideal sources for this goal.

MCG +8-11-11 ($z = 0.0204$) is a very X-ray bright AGN with a black hole mass of $\log \frac{M_{BH}}{M_\odot} = 7.19 \pm 0.02$ (Bian & Zhao 2003) and X-ray fluxes, measured by *INTEGRAL*, of $F_{20-100\text{keV}} = 8.46 \times 10^{-11}$ erg cm$^{-2}$ s$^{-1}$ and $F_{2-10\text{keV}} = 5.62 \times 10^{-11}$ erg cm$^{-2}$ s$^{-1}$ (Malizia et al. 2012). The $2 - 10$ keV absorption-corrected luminosity of the source is $6.45 \pm 0.04 \times 10^{43}$ erg s$^{-1}$ (Bianchi et al. 2010). *ASCA* (Grandi et al. 1998) and BeppoSAX (Perola et al. 2000) showed that the spectrum is well fitted by a model composed by a power law, a warm absorber, a Compton reflection component and an FeK$\alpha$ line. The best fit of the *ASCA* and OSSE data was an absorbed power law with spectral index $\Gamma = 1.73 \pm 0.06$ and an exponential cutoff at ~250 keV, plus a reflection component and a cold iron line. Also BeppoSAX data showed the presence of a cutoff at high energy (~ 170 keV). The XMM-*Newton* spectrum (Matt et al. 2006) revealed the lack of a soft excess, a large reflection component and a narrow iron line with a low equivalent width (EW) and no relativistic features. Bianchi et al. (2010) found in the *Suzaku* observation a relativistic Fe K$\alpha$ line, plus a narrow component with no associated reflection continuum.

NGC 6814 ($z = 0.0052$, Molina et al. 2009) is a Seyfert 1 Galaxy with black hole mass of $\log \frac{M_{BH}}{M_\odot} = 6.99^{+0.32}_{-0.25}$ (Pancoast et al. 2014, 2015) known to show X-ray variability by at least a factor of 10 over time scales of years (Mukai et al. 2003). It is part of the reverberation mapping campaign "the LAMP project" (Lick AGN Monitoring Project Bentz et al. 2009). The hard and soft X-ray flux of this source is $F_{20-100\text{keV}} = 5.66 \times 10^{-11}$ erg cm$^{-2}$ s$^{-1}$ and $F_{2-10\text{keV}} = 0.17 \times 10^{-11}$ erg cm$^{-2}$ s$^{-1}$ (Malizia et al. 2012).The INTEGRAL spectrum (Malizia et al. 2014) showed that the source has a quite flat spectrum ($\Gamma = 1.68 \pm 0.02$) with an exponential cut-off at $E_c = 190^{+185}_{-66}$ keV. From the XMM-*Newton* observation it is possible to see the presence of a narrow FeK$\alpha$ line (Ricci et al. 2014) with EW=$82^{+17}_{-15}$ eV. The *Suzaku* observation shows significant variability, a primary continuum with a photon index of $1.53 \pm 0.02$ and no evidence of soft excess. It shows also the emission of the Fe K$\alpha$ and Fe XXVI lines with centroid energy and EW respectively $E_{\text{FeK}\alpha} = 6.40 \pm 0.03$ keV, EW$_{\text{FeK}\alpha} = 170^{+30}_{-40}$ eV and $E_{\text{FeXXVI}} = 6.94 \pm 0.07$ keV, EW$_{\text{FeXXVI}} = 90^{+30}_{-40}$ eV (Walton et al. 2013).

In this paper we present the *NuSTAR* observations of MGC +8-11-11 (100 ks) and of NGC 6814 (150 ks), taken almost simultaneously with short *Swift* observations (20 ks each). In Sect.2 the observations and data reduction are presented. In Sect.3 we report on the spectral analysis of the two sources. The results are discussed and summarized in Sect.4.

## 2 OBSERVATIONS & DATA REDUCTION

### 2.1 NuSTAR

MCG +8-11-11 and NGC 6814 were observed by *NuSTAR* with its two coaligned X-ray telescopes (Focal Plane Modules A and B) respectively on 2016, August 19, and on 2016, July, 04. No other sources apart from the targets are apparent in the images.The *NuSTAR* data were reduced with the *NuSTAR* Data Analysis Software (NuSTARDAS) package (v. 1.6.0). Cleaned event files (level 2 data products) were produced and calibrated using standard filtering criteria with the NUPIPELINE task using the last calibration files available from the *NuSTAR* calibration database (CALDB 20170120). The extraction radii of the circular region were 0.5 arcmin for source and 1.5 arcmin for background spectra; there are no other bright X-ray sources within 1.5 arcmin from MGC 8-11-11 and NGC 6814 and no other sources were present in the background region. Net exposure time, after this process, is 98 ks for MGC 8-11-11 and 148 ks for NGC 6814 for both FPMA and B. The spectra were binned in order to over-sample the instrumental resolution by at least a factor of 2.5 and to have a Signal-to-Noise Ratio (SNR) greater than 5 for both sources in each spectral channel.

## 2.2 Swift-XRT

MCG +8-11-11 and NGC 6814 were observed by *Swift* UVOT+XRT almost simultaneously with *NuSTAR* for a total exposure time of 20 ks each. *Swift* XRT spectra were extracted using the XSELECT (v2.4c) command line interface to the FTOOLS (Blackburn 1995).

If there is pile-up the measured rate of the source is high (above about 0.6 counts s$^{-1}$ in the Photon-Counting Mode). The easiest way to avoid problems related to pile-up is to extract spectra using an annular region, thus eliminating the counts in the bright core, where pile-up will occur.

The *Swift* XRT spectra resulted to have a high pile-up degree. We tested different annular extraction regions for the source, gradually increasing the inner radius. Even with large inner extraction radii, the pile-up was not removed and, moreover, the signal-to-noise ratio became very low. We, therefore, decided not to use the *Swift* XRT data for the spectroscopic analysis.

## 3 DATA ANALYSIS

The spectral analysis has been performed with the XSPEC 12.9.0 software package (Arnaud 1996). Throughout the paper, errors correspond to 90% confidence level for one interesting parameter ($\Delta \chi^2 = 2.7$), if not stated otherwise. The cosmological parameters $H_0 = 70$ km s$^{-1}$ Mpc$^{-1}$, $\Omega_\Lambda = 0.73$ and $\Omega_m = 0.27$, are adopted.

Both sources show variability in their light curves (especially NCG 6814, well known to be a variable source: Mukai et al. 2003, Walton et al. 2013). The variability of NGC 6814 was consistent with the softer-when-brighter behaviour and with the fact that its black hole mass is 1/10 times the black hole mass of MCG +8-11-11 but since no strong spectral variations were found in the ratio between the 10 − 80 and 3 − 10 keV count rates (see Figure 1) we decided to use time-averaged spectra for both sources.

We performed our data analysis by fitting the 3 − 80 keV *NuSTAR* spectra with different models, each of them including Galactic absorption with column densities $N_H = 1.84 \times 10^{21}$ cm$^{-2}$ for MCG +8-11-11 and $N_H = 9.11 \times 10^{20}$ cm$^{-2}$ for NGC 6814, as derived from HI maps (Kalberla et al. 2005). We tested the presence of additional intrinsic absorbers at the redshift of the sources, which in both cases resulted to be negligible in the *NuSTAR* band. The *NuSTAR* FPMA calibration constants are fixed to 1.0 while we left the *NuSTAR* FPMB cross-calibration constants free to vary. The values found for the constant for MGC +8-11-11 and NGC 6814 are respectively $1.034 \pm 0.009$ and $0.988 \pm 0.008$. These values are consistent with the expectation (Madsen et al. 2015).

### 3.1 X-ray/optical ratio

We used *Swift* data to compute the optical to X-ray spectral index ($\alpha_{ox}$), defined as:

$$\alpha_{ox} = -\frac{\log \left[ L_{2keV}/L_{2500Å} \right]}{2.607} \quad (1)$$

The $\alpha_{ox}$ index is the slope of a hypothetical power law between 2500Å and 2 keV rest-frame frequencies. The optical to X-ray ratio provides information about the balance between the accretion disk and the corona. The $\alpha_{ox}$ is found to be strongly anti-correlated with the ultraviolet luminosity density per unit frequency (see Lusso et al. 2010, Vagnetti et al. 2013 and references therein). The observed $\alpha_{ox} - L_{2500Å}$ correlation implies that AGN redistribute their energy in the UV and X-ray bands depending on the overall luminosity; more optical luminous AGN emit less X-ray per unit UV luminosity than less luminous AGNs (Strateva et al. 2005).

The *Swift*/UVOT observations were analysed taking advantage of the on-line tool multi-mission archive at the Asi Science Data Center (ASDC) website[1]. Using this tool we performed an on-line interactive analysis for all the available observations for the sources MCG+8-11-11 and NGC 6814, 3 and 4 respectively. This on-line tool runs the standard UVOT pipeline and generates a sky map of the observation in the selected available filter. When the source is detected it is possible to select it and run the UVOT aperture photometry. Using this tool it is possible to compute the monochromatic flux in the selected filter. For all the observations we extract a circular region for the source with a radius of 5 arcsec and a properly selected annular region for the background. We obtain monochromatic and extinction-corrected fluxes for MCG+8-11-11 at 3465Å (U), 2246Å (UVM2), 1928Å (UVW2) and for NGC 6814 for the same filters and UVW1 (2600Å). Flux at 2500Å was then computed interpolating the monochromatic flux measures for both the sources. We obtained $F_{2500Å} = 3.98 \times 10^{-26}$ erg cm$^{-2}$ s$^{-1}$ Hz$^{-1}$ and $F_{2500Å} = 4.36 \times 10^{-26}$ erg cm$^{-2}$ s$^{-1}$ Hz$^{-1}$ respectively for MCG +8-11-11 and NGC 6814.

The 2 keV fluxes are extrapolated from the NuSTAR data. We found $F_{2keV} = 5.25 \pm 0.05 \times 10^{-28}$ erg cm$^{-2}$ s$^{-1}$ Hz$^{-1}$ for MCG +8-11-11 and $F_{2keV} = 3.42 \pm 0.05 \times 10^{-29}$ erg cm$^{-2}$ s$^{-1}$ Hz$^{-1}$ for NGC 6814.

With these values we computed the $\alpha_{ox}$ using the equation (1) obtaining 1.11 for MCG +8-11-11 and 1.36 for NGC 6814.

We use the values of $L_{2500Å}$, obtained from the previous values of the fluxes, to compute the X-ray/optical ratio with the relation found by Lusso et al. (2010):

$$\alpha_{ox}(L_{2500Å}) = (0.154 \pm 0.0010) \log L_{2500Å} - (3.176 \pm 0.233) \quad (2)$$

to check that the analysed sources follow the trend of the sources analysed by Lusso et al. (2010). We found $\alpha_{ox} = 1.23 \pm 0.32$ and $\alpha_{ox} = 1.51 \pm 0.25$ respectively for MCG +8-11-11 and NGC 6814. These values are consistent, with the values of $\alpha_{ox}$ we found before. Moreover the sources follow also the trend of the $L_X - L_{UV}$ of Lusso & Risaliti (2017).

### 3.2 MCG +8-11-11

We started the spectral analysis by fitting the 3 − 80 keV *NuSTAR* spectrum with a phenomenological baseline model composed of a power law with an exponential cutoff for the primary continuum (CUTOFFPL), a Gaussian line to reproduce the narrow Fe K$\alpha$ emission line at 6.4 keV, and a cold reflection component (PEXRAV. Magdziarz & Zdziarski 1995). We fixed all element abundances to Solar values and the inclination angle to $\cos i = 0.86$, $i (\sim 30°)$. Since we know from the literature the presence of an emission line from H-like Fe K$\alpha$ (Bianchi et al. 2010), we added another narrow Gaussian line, with the centroid energy fixed to 6.966 keV. We found a $\chi^2 = 467$ for 427 degrees of freedom (d.o.f.). The residuals around the narrow Fe K$\alpha$ line suggest the presence of a broad line component. We therefore added a broad Gaussian line (hereafter model A1) to the previous model. The fit slightly improved, $\chi^2$/d.o.f = 446/424. We found a resolved Fe K$\alpha$ line with $\sigma = 0.31^{+0.15}_{-0.20}$ keV and the centroid energy at $6.21^{+0.18}_{-0.28}$ keV. The fitting parameters for the three lines are shown in Table 1. We found a 3-80 keV flux of $1.45 \pm 0.03 \times 10^{-10}$ erg cm$^{-2}$ s$^{-1}$ Regarding the

---
[1] http://www.asdc.asi.it

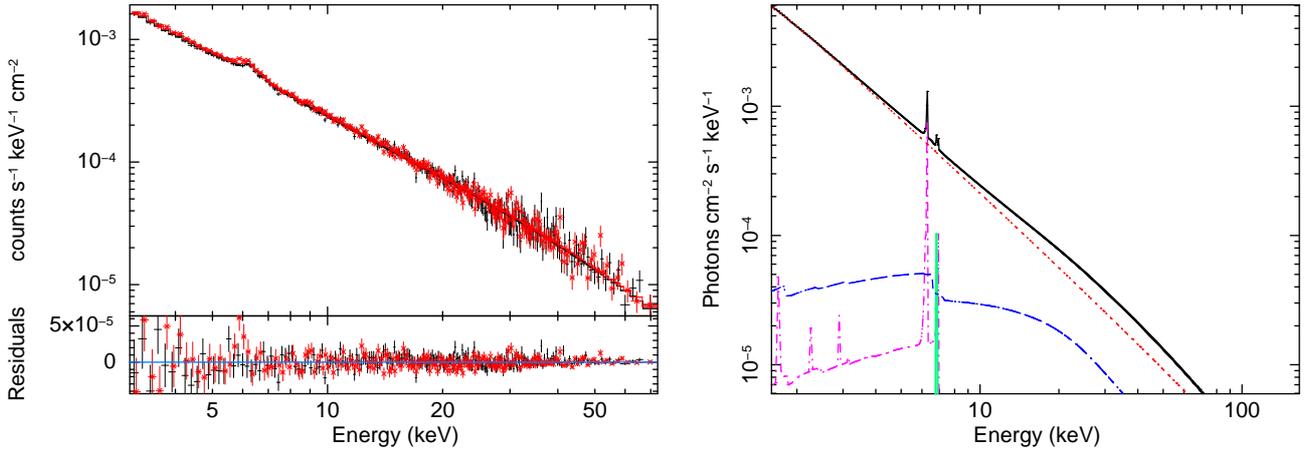

**Figure 2.** Left panel: data and best fit model of MCG +8-11-11 extrapolated from *NuSTAR* FPMA (black dot) and FPMB (red asterisk) spectra when model C1 (right panel) was used. Right panel: best fitting model (black solid line); the blue dashed line is the RELXILL component, the red dotted line represent the primary continuum, the bold solid green line represent the Fe XXVI emission line and the magenta dash-dotted line is the XILLVER component. See text for more details.

**Table 1.** Fit parameters for the emission features of MCG +8-11-11. Errors are at 90% confidence levels.

| Parameter | Fe K$\alpha$ NL | Fe K$\alpha$ BL | Fe XXVI |
|---|---|---|---|
| E (keV) | $6.40^{+0.03}_{-0.06}$ | $6.21^{+0.18}_{-0.28}$ | $6.966^\star$ |
| $\sigma$ (keV) | $0.0^\star$ | $0.31^{+0.15}_{-0.20}$ | $0.0^\star$ |
| EW (eV) | $40 \pm 15$ | $90 \pm 7$ | $28^{+3}_{-7}$ |
| Flux ($10^{-5}$ ph/cm$^2$/s) | $2.56^{+2.5}_{-3.7}$ | $5.78 \pm 0.9$ | $1.4 \pm 0.3$ |

$\star$ fixed parameter.

continuum, we found a power law index of $1.77 \pm 0.02$ and a cutoff energy of $260^{+190}_{-80}$ keV. The reflection fraction R was $0.15 \pm 0.06$ (Table 2, model A1).

The following step of the analysis was to replace the PEXRAV model plus the broad line with the RELXILL model (García et al. 2014), in order to test for the presence of a relativistic component (hereafter model B1). The inclination angle was fixed to a value of $30°$ and we fixed the ionization parameter to $\log \left( \frac{\xi_i}{\text{erg cm s}^{-1}} \right) = 0.0$ to test reflection from neutral material. Leaving the iron abundance free to vary we found a value of iron abundance of $A_{\text{Fe}} = 3.1^{+1.7}_{-1.4}$, a reflection fraction R= $0.24^{+0.12}_{-0.07}$ and a lower limit on the spin of the central black hole of $a > 0.6$. The $\chi^2/\text{d.o.f.}$ for this fit was 452/425. Other fitting parameters are shown in Table 2.

Leaving the ionization parameter free to vary the fit did not improve, the $\chi^2/\text{d.o.f.}$ remained the same, and we found an upper limit value for the ionization parameter of $\log \left( \frac{\xi_i}{\text{erg cm s}^{-1}} \right) < 0.05$.

We then substituted the Gaussian narrow line in the B1 model with the XILLVER model (García & Kallman 2010; García et al. 2013) (hereafter model C1, see right panel of Figure 2), to test if the reprocessed spectrum could originate in distant material, like a Compton-thick torus. The photon indices the cutoff energy of the XILLVER and RELXILL models were tied together. The reflection fraction of XILLVER was free to vary. The fit gave a $\chi^2/\text{d.o.f.}= 457/425$. We found a value for the high energy cutoff of $175^{+110}_{-50}$ keV (see Table 2). With this model the observed reflection component appeared to be mostly associated to the accretion disk and to the broad part of the Fe K$\alpha$. The narrow part of the Fe K$\alpha$ line had an emission which was relatively strong compared to the Compton hump, and the reflection fraction of the XILLVER component was found to be quite low: $R^{\text{xill}} = 0.25 \pm 0.12$. This may be due to emission from a material with high iron abundance, and indeed we found an iron overabundance in the XILLVER model of $A_{\text{Fe}} > 8.5$. The other fit parameters are reported in Table 2. Contour plots are shown in Figure 4.

If we tried to keep tied the iron abundances between the XILLVER and the RELXILL model we found a value of $3.2 \pm 0.7$ but the fit worsened significantly with a resulting $\chi^2/\text{d.o.f.}= 514/425$.

We tested if the large ratio between the line flux and the Compton hump may be due to reflection from a distant material with $N_H < 10^{24}$cm$^{-2}$ using the MYTORUS model (Murphy & Yaqoob 2009). This is a more physical model respect to the phenomenological model C1, that we used to have a measure of the cutoff. We used the MYTORUS model to fit the cold reflection and the Fe K$\alpha$ and Fe K$\beta$ emission lines (and the associated Compton shoulders) adding the MYTORUS Compton-scattered continuum and fluorescent line tables to the RELXILL component (hereafter model D1). In the fit, we kept tied the photon indices of the MYTORUS table to the RELXILL one. The iron abundance for the RELXILL model was fixed to the values of the model C1, while for MYTORUS it is not a variable parameter being the Solar value. In the standard "coupled" configuration the inclination angle for the torus was fixed to $\theta = 30°$. We obtained a $\chi^2/\text{d.o.f.}= 445/435$. The best-fit parameters are given in Table 2. We found a column density of $N_H = 5.0 \pm 3.0 \times 10^{23}$cm$^{-2}$, and only a lower limit to the cutoff energy of $E_c > 250$ keV. It must be remarked that the model is not fully self-consistent in this respect because in MYTORUS the illuminating continuum must be a straight power law, with a sharp terminal energy of $\sim 500$ keV. However, the lower limit of the cutoff energy is consistent with the values found with the previous models. Moreover, the reflection from distant, Compton thin material is in agreement with what we found in model C1, where the relativistic reflection dominates the spectral shape above 10 keV, with respect to the non-relativistic one.

Finally, we tried to see if the two components, modelled by XILLVER and RELXILL, are not two different physical component (one arising from the accretion disk and one from a distant material: a Compton-thin gas or a gas with a super-solar abundance of iron) but two different reflection components arising from different parts of the same region, in this case the accretion disk which is in different configuration. We modeled the spectrum with two relativistic components (two RELXILL models). Keeping all the parameters tied tighter between the two model, except for the normalization, the resulting chi-square of the fit is $\chi^2$/d.o.f.= 486/426 = 1.13 . We found a lower limit on the iron abundance of $A_{Fe} > 5.88$ and there are also clear residuals around 6.4 keV which indicate the presence of the narrow component of the Fe K$\alpha$ line. Adding this component and allowing to vary the iron abundance parameter, the reflection fraction parameter, and the black hole spin parameter, we found a fit which is statistically equivalent to our best fit model, with a chi-square of $\chi^2$/d.o.f.= 452/424 = 1.07. We found one RELXILL component with an upper limit on the iron abundance of $A_{Fe} < 1.48$ an upper limit on the spin parameter of $a < 0.48$ and a reflection fraction R= $0.16 \pm 0.05$. The second relxill component showed a lower limit on the iron abundance of $A_{Fe} > 6.35$, a spin value of $a = 0.07 \pm 0.01$ and a reflection fraction R< 0.07. This suggested us that the latter is not a relativistic component and it could not be produced by a Compton thick material since it had a very low associated reflection component and high iron abundance, similarly to the XILLVER component of our best fit (model C1). This scenario is consistent with what Bianchi et al. (2010) and Mantovani et al. (2016) found for this source.

**Table 2.** Fitting parameters for MCG +8-11-11 for the models A1, B1, C1, D1. Errors are at 90% confidence levels. In the range of 3-0 keV we found a flux of $1.45 \pm 0.03 \times 10^{-10}$ erg cm$^{-2}$ s$^{-1}$ and an absorption corrected luminosity of $1.37 \pm 0.05 \times 10^{44}$ erg s$^{-1}$.

| Model: | PEXRAV (A1) | RELXILL (B1) | RELXILL+XILLVER (C1) | RELXILL+MYTORUS (D1) |
|---|---|---|---|---|
| $\Gamma$ | $1.77 \pm 0.02$ | $1.77 \pm 0.03$ | $1.77 \pm 0.04$ | $1.83 \pm 0.03$ |
| $E_c$ (keV) | $260^{+190}_{-80}$ | $224^{+140}_{-70}$ | $175^{+110}_{-50}$ | $> 250$ |
| $R^{PEXRAV}$ | $0.15 \pm 0.06$ | - | - | - |
| $R^{RELXILL}$ | - | $0.24^{+0.12}_{-0.07}$ | $0.25 \pm 0.12$ | $0.23 \pm 0.08$ |
| $R^{XILLVER}$ | - | - | $0.17 \pm 0.04$ | - |
| $a$ | - | $> 0.6$ | $> 0.5$ | $< 0.72$ |
| $A_{Fe}^{RELXILL}$ (Solar Units) | - | $3.1^{+1.7}_{-1.4}$ | $< 2.5$ | $2.5^\star$ |
| $A_{Fe}^{XILLVER}$ (Solar Units) | - | - | $> 8.5$ | - |
| $N_H^{MYT}$ ($10^{23}$cm$^{-2}$) | - | - | - | $5.0 \pm 3.0$ |
| $\frac{\chi^2}{d.o.f}$ | 1.05 | 1.06 | 1.07 | 1.05 |

★ fixed parameter.

Fluxes and centroid energies of the broad and narrow Fe K$\alpha$ lines and of the Fe XXVI line resulted to be the same, within the errors, among the various models previously described.

### 3.2.1 Coronal parameters

Assuming that the primary emission is due to Comptonization of thermal disc photons in a hot corona, we estimated the coronal parameters substituting the cut-off power law with a Comptonization model. We modelled the relativistic reflection and the non-relativistic reflection using respectively RELXILLCP and XILLVER-COMP models (García & Kallman 2010; García et al. 2014 and Dauser et al. 2014). These models use the NTHCOMP model (Zdziarski et al. 1996 and Życki et al. 1999) for the incident continuum. In RELXILLCP and XILLVER-COMP models, the maximum temperature of disc blackbody photons (which serve as seeds for Comptonization) is 0.1 keV, and it is not allowed to vary. The fit gave a $\chi^2$/d.o.f.= 465/426 = 1.09. We found a photon index value of $1.84^{+0.03}_{-0.05}$; we found also a lower limit for the iron abundance of the non-relativistic reflection component: $A_{Fe} > 8.01$ Solar Unit, and an upper limit for the iron abundance of the relativistic reflection component: $A_{Fe} < 0.5$ Solar Unit, in agreement with the values reported in Table 2. We found a coronal temperature of $60^{+110}_{-30}$ keV, roughly in agreement with the expected relation $E_c = 2 - 3 \times kT_e$ (Petrucci et al. 2000, 2001).

### 3.3 NGC 6814

The fitting procedure for NGC 6814 was very similar to that adopted for MCG +8-11-11. We started the analysis by fitting the 3 − 80 keV *NuSTAR* spectrum with the phenomenological baseline model, previously described in subsection 3.2 and composed by a cutoff power law, a PEXRAV plus a narrow Gaussian line around 6.4 keV. Also in this case we fixed all element abundances to Solar values and the inclination angle to $\cos i = 0.86$ ($i \sim 30°$). Since Walton et al. (2013) found the presence of an emission line from H-like Fe K$\alpha$ with an EW of $90^{+30}_{-40}$ eV, we added also a narrow Gaussian line with the centroid energy fixed to 6.966 keV. However, only an upper limit to the EW of < 15 eV was found, which was more than four times lower than the value found by Walton et al. (2013). We found also an upper limit for the line flux of $< 2.1 \times 10^{-6}$ ph/cm$^2$. This was probably due to the fact that the line was diluted by a continuum which was about four times higher than that found by Walton et al. (2013) (see below). We therefore did not include this line in this and in the following fits. We found a $\chi^2$/d.o.f.= 398/348. The residuals around the narrow Fe K$\alpha$ line suggested the presence of a broad line. Adding a broad Gaussian line at the PEXRAV plus the narrow Fe K$\alpha$ line model (model A2) around 6.4 keV we found a resolved Fe K$\alpha$ line with $\sigma = 0.59^{+0.37}_{-0.21}$ keV, EW of $102 \pm 15$ eV and flux of $4.2 \pm 1.4 \times 10^{-5}$ ph/cm$^2$/s. The centroid energy of the Fe K$\alpha$ line was found to be at $6.43^{+0.03}_{-0.06}$ keV. The line had an EW of $70 \pm 7$ eV, consistent with what Walton et al. (2013) found. The flux of the line was $2.7 \pm 0.7 \times 10^{-5}$ ph/cm$^2$/s. The chi square for the fit with the model A2 was $\chi^2$/d.o.f.= 373/345. The high energy cutoff was $260^{+220}_{-80}$ keV and the reflection fraction R= $0.15 \pm 0.07$. The 3 − 80 keV flux was $1.04 \pm 0.04 \times 10^{-10}$ erg cm$^{-2}$ s$^{-1}$. Other parameters are reported in Table 3.

The following step in the analysis was to replace the PEXRAV model plus the broad line with the RELXILL model, as for the previous source but without the Fe XXVI line (hereafter model B2) in order to test for the presence of a relativistic component. The inclination angle was fixed to 30° and the ionization parameter was fixed to

**Table 3.** Fitting parameters for NGC 6814 using the models A2, B2, C2, D2, as described in the text. Errors are at 90% confidence levels. In the range of 3-0 keV we found a flux of $1.04 \pm 0.04 \times 10^{-10}$ erg cm$^{-2}$ s$^{-1}$ and an absorption corrected luminosity of $6.21 \pm 0.12 \times 10^{42}$ erg s$^{-1}$.

| Model: | PEXRAV (A2) | RELXILL (B2) | RELXILL+XILLVER (C2) | RELXILL+MYTORUS (D2) |
|---|---|---|---|---|
| $\Gamma$ | $1.72 \pm 0.02$ | $1.74 \pm 0.02$ | $1.71^{+0.04}_{-0.03}$ | $1.80 \pm 0.02$ |
| $E_c$ (keV) | $260^{+220}_{-80}$ | $210^{+80}_{-50}$ | $155^{+70}_{-35}$ | $> 260$ |
| $R^{\mathrm{PEXRAV}}$ | $0.15 \pm 0.07$ | - | - | - |
| $R^{\mathrm{RELXILL}}$ | - | $0.26 \pm 0.1$ | $0.27^{+0.10}_{-0.12}$ | $0.26^{+0.18}_{-0.11}$ |
| $R^{\mathrm{XILLVER}}$ | - | - | $0.17 \pm 0.03$ | - |
| $a$ | - | $> 0.2$ | $> 0.03$ | $> 0.4$ |
| $A_{\mathrm{Fe}}^{\mathrm{RELXILL}}$ (Solar Units) | - | $< 1.4$ | $1.8^{+0.8}_{-1.3}$ | $1.8^\star$ |
| $A_{\mathrm{Fe}}^{\mathrm{XILLVER}}$ (Solar Units) | - | - | $> 7.00$ | - |
| $N_H^{\mathrm{MYT}}$ ($10^{23}$cm$^{-2}$) | - | - | - | $3.5^{+3.0}_{-2.0}$ |
| $\frac{\chi^2}{\mathrm{d.o.f}}$ | 1.08 | 1.06 | 1.08 | 1.06 |

$\star$ fixed parameter.

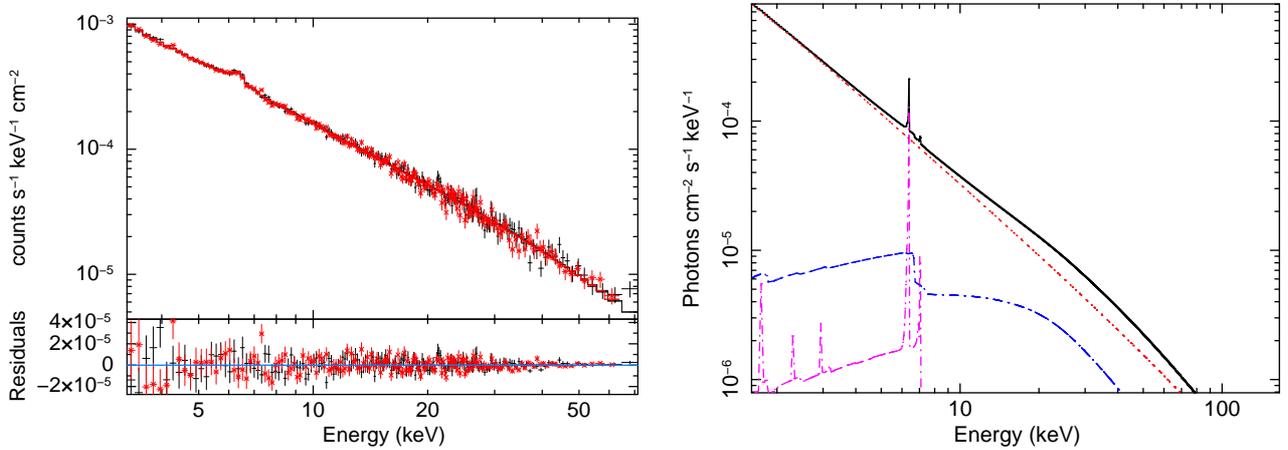

**Figure 3.** Left panel: data and best fit model of NGC 6814 extrapolated from *NuSTAR* FPMA (black dot) and FPMB (red asterisk) spectra when model C2 (right panel) is used. Right panel: best fitting model (solid black line); the blue dashed line is the RELXILL component, the red dotted line represent the primary continuum and the magenta dash-dotted line is the XILLVER component. See text for more details.

$\log\left(\frac{\xi_i}{\mathrm{erg\ cm\ s}^{-1}}\right) = 0.0$. Leaving the iron abundance free to vary we found a value of $A_{Fe} < 1.4$ in Solar Units, a reflection fraction R= $0.26 \pm 0.1$ and a lower limit on the spin of the central black hole of $a > 0.2$. The $\chi^2$ values in this case was $\chi^2$/d.o.f.= 366/343. Leaving the ionization parameter free to vary the fit did not improve, the $\chi^2$ remained the same, and we found an upper limit value for the ionization parameter of $\log\left(\frac{\xi_i}{\mathrm{erg\ cm\ s}^{-1}}\right) < 0.34$. Other fitting parameters are shown in Table 3.

We then replaced the Gaussian narrow line in the model B2 with the XILLVER model (hereafter model C2; see right panel of Figure 3), to test if the reprocessed spectrum could originate in a distant material, as it was done for the other source in the previous section. The photon indices the cutoff energy of the XILLVER and RELXILL models are tied together. The reflection fraction of XILLVER was free to vary. The fit gave a $\chi^2$/d.o.f.= 375/346. We found a value for the high energy cutoff of $155^{+70}_{-35}$ keV (see Table 3). The other fit parameters are reported in Table 3. Contour plots are shown in Figure 4. Also for NGC 6814, as for MCG +8-11-11, we found an iron overabundance in the XILLVER model, $A_{Fe} > 7.0$ due to an emission of the narrow part of the Fe K$\alpha$ line which was relatively strong compared to the Compton hump. If we tried to keep tied the iron abundances between the XILLVER and the RELXILL model we found a lower limit value of $A_{Fe} > 4$. and the fit worsened significantly with a resulting $\chi^2$/d.o.f.= 379/347.

Similarly to the case of MCG +8-11-11, we tested the alternative model in which XILLVER was replaced by reflection from distant material with $N_H < 10^{24}$cm$^{-2}$ that could reproduce the narrow part of the fluorescence emission line from the iron K-shell with a small Compton hump. We used the MYTORUS model as described in subsection 3.2 (hereafter model D2). We obtained a $\chi^2$/d.o.f.= 368/347. The best fit parameters are given in Table 3. We found a column density of $N_H = 3.5^{+3.0}_{-2.0} \times 10^{23}$cm$^{-2}$. Again, we found only a lower limit to the cutoff energy, $E_c > 260$ keV, which was consistent with the values found with the previous models.

Yamauchi et al. (1992) found an iron overabundance in the *Ginga* spectrum of the source. They justified the fact with the presence of a partially ionized state that may give an "apparent" overabundance because the partially ionized gas is transparent for the soft X-rays but absorbs X-rays above the Fe K-edge energy. In our fit, we considered neutral reflection material, so the super-solar value for the iron abundance could not be an effect of the ionization. Even in this case, we tested if the-the two reflection components could arise from two different parts of the accretion disk if it is in a different configuration. We modeled the spectrum with two RELXILL models. First, we kept all the parameters tied tighter between the two model, except for the normalization, the resulting chi-square of the fit is $\chi^2$/d.o.f.= 409/347 = 1.17. We found a lower limit on the iron abundance of $A_{Fe} > 5.88$ and there are also clear residuals around 6.4 keV which indicate the presence of the narrow component of the Fe K$\alpha$ line. Adding this component and allowing to vary the iron abundance parameter, the reflection fraction parameter, and the black hole spin parameter, we found a fit with a chi-square of $\chi^2$/d.o.f.= 373/343 = 1.09. We found one RELXILL component with an upper limit on the iron abundance of $A_{Fe} < 2.09$ an upper limit on the spin parameter of $a < 0.48$ and a reflection fraction R= $0.27 \pm 0.12$. The second relxill component showed a lower limit on the iron abundance of $A_{Fe} > 6.35$, an upper limit on the spin value $a < 0.15$ and a lower limit on the reflection fraction R< 0.12. Also NGC 6814 shows one of the RELXILL component which is not a relativistic component and it could not be produced by a Compton thick material since it had a very low associated reflection component and high iron abundance, similarly to the XILLVER component of our best fit (model C2). Fluxes of the broad and narrow Fe K$\alpha$ lines and of the Fe XXVI line were the same, within the errors, among the various models previously described.

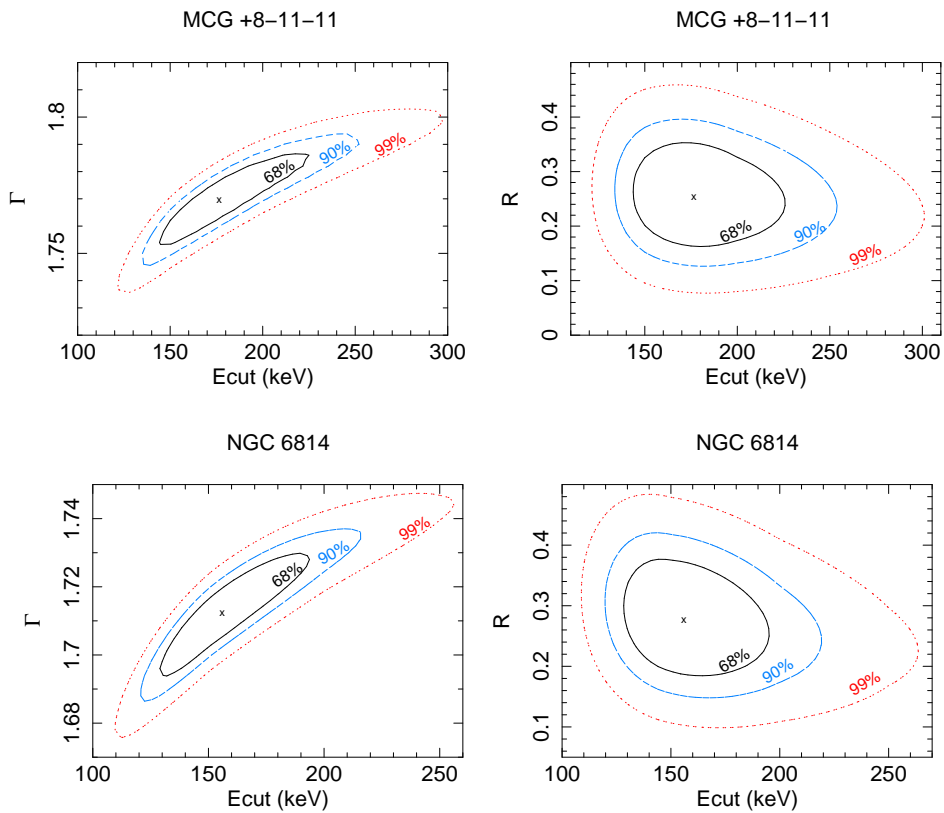

**Figure 4.** $E_c$-$\Gamma$ contour plot (left panels) and $E_c$-R contour plot (right panels) for MCG +8-11-11 (top panels) and NGC 6814 (lower panels). The black solid, blue dashed and red dotted curves refer to the 68, 90 and 99% confidence levels respectively. The X represents the best fit value of the parameters.

*3.3.1 Coronal parameters*

The final step was to use Comptonization models to estimate the coronal parameters as in section 3.2.1. We used RELXILLCP and XILLVER-COMP to model both the relativistic and non-relativistic reflection spectrum, see Section 3.2.1. The fit gave a $\chi^2$/d.o.f.= 387/347 = 1.11. We found a photon index value of $1.79 \pm 0.03$ and a coronal temperature of $45^{+100}_{-17}$ keV, again in agreement with the $E_c = 2-3 \times kT_e$ relation (Petrucci et al. 2000, 2001). We found also a lower limit for the iron abundance of the non-relativistic reflection component: $A_{Fe} > 7.89$ Solar Unit, and an upper limit for the iron abundance of the relativistic reflection component: $A_{Fe} < 0.73$ Solar Unit, in agreement with the values reported in Table 3.

**Table 4.** Coronal parameters for MCG +8-11-11 and NGC 6814 when the self-consistent model XILLVER-COMP + RELXILLCP is used to fit the data. The optical depths are extrapolated from Beloborodov (1999). Errors are at 90% confidence levels.

| Parameter | MCG +8-11-11 | NGC 6814 |
|---|---|---|
| $kT_e$ (keV) | $60^{+110}_{-30}$ | $45^{+100}_{-17}$ |
| $\tau$ | $1.8 \pm 0.2$ | $2.5 \pm 0.2$ |
| $\Gamma$ | $1.84^{+0.03}_{-0.05}$ | $1.79 \pm 0.03$ |
| $A_{Fe}^{RELXILL}$ (Solar Units) | < 0.5 | < 0.73 |
| $A_{Fe}^{XILLVER}$ (Solar Units) | > 8.01 | > 7.89 |
| $\frac{\chi^2}{d.o.f}$ | 1.09 | 1.11 |

## 4 DISCUSSION AND CONCLUSIONS

We have presented the analysis of the *NuSTAR* observations of the Seyfert 1 galaxies NGC 6814 and MCG +8-11-11.

The 2 − 10 keV absorption-corrected luminosities from the *NuSTAR* observations of the two sources are $L_{2-10} = 2.04 \times 10^{42}$ erg s$^{-1}$ for NGC 6814 and $L_{2-10} = 5.13 \times 10^{43}$ erg s$^{-1}$ for MCG +8-11-11. Using the 2 − 10 keV bolometric correction of Marconi et al.

(2004), we estimated the bolometric luminosity to be $L_{bol} = 0.24 \times 10^{44}$ erg s$^{-1}$ (NGC 6814) and $L_{bol} = 14.2 \times 10^{44}$ erg s$^{-1}$ (MCG +8-11-11). From these bolometric luminosity, with the black hole masses of $\log \frac{M_{BH}}{M_\odot} = 7.19 \pm 0.02$ (Bian & Zhao 2003) (MCG +8-11-11) and $\log \frac{M_{BH}}{M_\odot} = 6.99^{+0.32}_{-0.25}$ (Pancoast et al. 2014, 2015) (NGC 6814), we estimated the Eddington ratio to be $2.46 \times 10^{-3}$ for NGC 6814 and $7.54 \times 10^{-1}$ for MGC +8-11-11.

Thanks to the *NuSTAR* sensitivity at high energies, it was possible to measure the high energy cutoff value for both sources; we found $175^{+110}_{-50}$ keV and $155^{+70}_{-35}$ respectively for MCG +8-11-11 and NGC 6814. We found also a disk reflection component and we constrained the reflection fraction founding $0.25 \pm 12$ for MCG +8-11-11 and of $0.27^{+0.10}_{-0.12}$ for NGC 6814.

Both sources showed a slightly broadened relativistic Fe K$\alpha$ line plus a narrow component. The reflection component was modest, and mostly associated with the broad line component. The low reflection fraction found in MCG +8-11-11 was consistent with the value found by Mantovani et al. (2016). Past observations of NGC 6814 did not show broad line ( Bentz et al. 2009, Malizia et al. 2014, Ricci et al. 2014) . The Compton hump associated with the narrow line was very small, similarly to what found in another Seyfert galaxy, NGC 7213 (Ursini et al. 2015).

We found in our analysis a slightly broadened relativistic Fe K$\alpha$ line plus a narrow component. The former is modelled by the relativistic model RELXILL while the latter by the non-relativistic model XILLVER. The iron abundance measured with the RELXILL model is considerably lower with respect to the super-Solar abundance measured with the XILLVER model. Since there was no Compton reflection hump associated with the narrow component of the iron K$\alpha$ line, it could not be produced by a Compton-thick material, like the accretion disc or the Compton-thick torus, thus almost all the reflection should be associated with the accretion disk.
The interaction of X-rays with a material with super-Solar abundance of iron, gives rise to a reflection component with small Compton hump associated with the narrow iron K$\alpha$ emission line. This is because the iron atoms interact with the X-rays, causing photoelectric absorption and the spectrum shows an iron K$\alpha$ line with a drop at $\sim 7$ keV, due to the photoelectric absorption edge. The depth of the edge increases with the iron abundance and saturates around $A_{Fe} \sim 10$ (Matt et al. 1997). The curvature of the continuum above $\sim 10$ becomes weaker, resulting in a very small Compton reflection. A reflection component with small Compton hump associated with the narrow iron K$\alpha$ emission line could be produced also by the interaction of X-rays with a Compton thin material. To confirm this scenario we fitted the cold reflection component with the MYTORUS model and we found a values of $N_H < 10^{24}$ cm$^{-2}$. This could implies that there would be no evidence of the classical Compton-thick torus in these sources, as already suggested by Bianchi et al. (2010) for MCG +8-11-11.

Ultimately we tested also if the two different reflection components should be part of the emission by the same material, such as an accretion disk which is in different configuration. But this attempt has further strengthened the scenario described above.

Regarding the relativistic component, for both sources we derived only lower or upper limits to the spin of the black hole of the two sources, depending on the adopted model. The relativistic line was fairly broad, so there are degeneracies in the parameters of the models which prevent us from determining a good measurement of the spin. We conclude that this parameter is basically unconstrained, not surprisingly given the relative weakness of the relativistic reflection.

We estimated the coronal parameters by fitting the *NuSTAR* data with a model which takes into account both the relativistic and non-relativistic reflection when illuminated by a thermally Comptonized continuum. We found a coronal temperature of $60^{+110}_{-30}$ keV for MCG +8-11-11 and $45^{+100}_{-17}$ keV for NGC 6814. It is interesting to note that the coronal temperature of the two sources are very similar despite an order of magnitude difference in mass and Eddington ratio.

We estimated the optical depth using the relation from Beloborodov (1999):

$$\Gamma \approx \frac{9}{4} y^{-2/9} \quad (3)$$

where $\Gamma$ is the photon index of the spectrum between 2 and 10 keV. The dependence from the optical depth is in the relativistic *y*-parameter:

$$y = 4 \left( \Theta_e + 4\Theta_e^2 \right) \tau(\tau + 1) \quad (4)$$

where $\Theta_e$ is the electron temperature normalized to the electron rest energy:

$$\Theta_e = \frac{kT_e}{m_e c^2} \quad (5)$$

We found $\tau = 1.79 \pm 0.2$ for MCG +8-11-11 and $\tau = 2.5 \pm 0.2$ for NGC 6814.

We used the values of the coronal temperature reported in Table 4 to put MCG +8-11-11 and NGC 6814 in the compactness-temperature ($\Theta_e$ - $\ell$) diagram (Fabian et al. 2015, and references therein). Here $\Theta_e$ is the electron temperature normalized to the electron rest energy, defined above in equation 5, and $\ell$ is the dimensionless compactness parameter (Fabian et al. 2015):

$$\ell = \frac{L}{R} \frac{\sigma_T}{m_e c^3} \quad (6)$$

where L is the luminosity and R is the radius of the corona (assumed spherical). To compute the compactness parameter, following Fabian et al. (2015) we adopted the luminosity of the power-law component extrapolated to the $0.1 - 200$ keV band; since no measurement exists for the corona radius, we assume a value of 10 gravitational radii $R_g$. For MCG +8-11-11 we found $\ell = 27 \pm 12(R_{10})^{-1}$ and $\Theta_e = 0.11^{+0.15}_{-0.10}$; for NGC 6814 $\ell = 14.5 \pm 4.5(R_{10})^{-1}$ and $\Theta_e = 0.08^{+0.1}_{-0.05}$ . Here $R_{10}$ is the ratio between the corona radius and $10R_g$.

With these values of $\Theta_e$ and $\ell$ both sources are positioned under the Svensson pair runaway line for a spherical geometry (Svensson 1984), above the $e^- - e^-$ coupling line, like most of the sources among those analysed by Fabian et al. (2017). The pair runaway line in the $\Theta_e$ - $\ell$ diagram is a curve which determine a forbidden region in which the pair production exceeds the annihilation. The detailed shape of this line depends on the source of soft photons and on the radiation mechanism. Above the $e^- - e^-$ coupling line the $e^- - e^-$ coupling time scale is longer than the Compton cooling time scale. The location of these sources within the $\Theta_e$ - $\ell$ plane fits well in the scenario in which the AGN spectral shape is controlled by pair production and annihilation.

**ACKNOWLEDGEMENTS**

We thank the anonymous referee for comments that helped improving the clarity of the paper. This work made use of data from the *NuSTAR* mission, a project led by the California Institute of Technology, managed by the Jet Propulsion Laboratory, and funded by the National Aeronautics and Space Administration. We thank the *NuSTAR* Operations, Software and Calibration teams for support with the execution and analysis of these observations. This research has made use of the *NuSTAR* Data Analysis Software (NuSTAR-DAS) jointly developed by the ASI Science Data Center (ASDC, Italy) and the California Institute of Technology (USA). AT, AM and GM acknowledge financial support from Italian Space Agency under grant ASI/INAF I/037/12/0-011/13, SB under grant ASI-INAF I/037/12/P1. Part of this work is based on archival data, software or on-line services provided by the ASI SCIENCE DATA CENTER (ASDC). AT, SB, AM,GM and RM acknowledge financial support from the European Union Seventh Framework Programme (FP7/2007-2013) under grant agreement n.312789. POP acknowledges financial support from the CNES and the CNRS/PNHE. EP and SB acknowledge financial contribution from the agreement ASI-INAF I/037/12/0.

This paper has been typeset from a TeX/LaTeX file prepared by the author.